\documentclass[pra,onecolumn,superscriptaddress]{revtex4-1}

\usepackage{amssymb}
\usepackage{graphicx}
\usepackage{bbm,amsmath,amssymb}
\usepackage{array}
\usepackage{caption}
\usepackage{subcaption}

\usepackage{graphicx}
\usepackage[usenames,dvipsnames]{color}
\usepackage{epsfig,color}
\usepackage{amsmath,amssymb}

\usepackage[T1]{fontenc}
\usepackage{verbatim}
\usepackage{float}
\usepackage{units}
\usepackage{amsmath}
\usepackage{graphicx}
\usepackage[colorlinks = true,
            linkcolor = NavyBlue,
            urlcolor  = NavyBlue,
            citecolor = NavyBlue,
            anchorcolor = blue]{hyperref}
\usepackage{epstopdf}

\makeatletter

 \@ifundefined{textcolor}{}
 {
   \definecolor{BLACK}{gray}{0}
   \definecolor{WHITE}{gray}{1}
   \definecolor{RED}{rgb}{1,0,0}
   \definecolor{GREEN}{rgb}{0,1,0}
   \definecolor{BLUE}{rgb}{0,0,1}
   \definecolor{CYAN}{cmyk}{1,0,0,0}
   \definecolor{MAGENTA}{cmyk}{0,1,0,0}
   \definecolor{YELLOW}{cmyk}{0,0,1,0}
 }

\makeatother

\def\laq{\lower 0.4ex\hbox{$<$}\kern -0.8em\raise 0.62
ex\hbox{$\sim$}}
\def\gaq{\lower 0.4ex\hbox{$>$}\kern -0.7em\raise 0.62
ex\hbox{$\sim$}}

\begin{document}
\title{Quantum Teleportation of Propagating Quantum Microwaves}

\author{R. Di Candia}
\email{rob.dicandia@gmail.com}
\affiliation{Department of Physical Chemistry, University of the Basque Country UPV/EHU, Apartado   644, 48080 Bilbao, Spain}
\author{K. G. Fedorov}
\affiliation{Walther-Mei\ss ner-Institut, Bayerische Akademie der Wissenschaften, D-85748 Garching, Germany}
\affiliation{Physik-Department, Technische Universit\"at M\"unchen, D-85748 Garching, Germany}
\author{L. Zhong}
\affiliation{Walther-Mei\ss ner-Institut, Bayerische Akademie der Wissenschaften, D-85748 Garching, Germany}
\affiliation{Physik-Department, Technische Universit\"at M\"unchen, D-85748 Garching, Germany}
\affiliation{Nanosystems Initiative Munich (NIM), Schellingstra\ss e 4, 80799 M\"unchen, Germany}
\author{S. Felicetti}
\affiliation{Department of Physical Chemistry, University of the Basque Country UPV/EHU, Apartado   644, 48080 Bilbao, Spain}
\author{\\E. P. Menzel}
\affiliation{Walther-Mei\ss ner-Institut, Bayerische Akademie der Wissenschaften, D-85748 Garching, Germany}
\affiliation{Physik-Department, Technische Universit\"at M\"unchen, D-85748 Garching, Germany}
\author{M. Sanz}
\affiliation{Department of Physical Chemistry, University of the Basque Country UPV/EHU, Apartado   644, 48080 Bilbao, Spain}
\author{F. Deppe}
\affiliation{Walther-Mei\ss ner-Institut, Bayerische Akademie der Wissenschaften, D-85748 Garching, Germany}
\affiliation{Physik-Department, Technische Universit\"at M\"unchen, D-85748 Garching, Germany}
\affiliation{Nanosystems Initiative Munich (NIM), Schellingstra\ss e 4, 80799 M\"unchen, Germany}
\author{A. Marx}
\affiliation{Walther-Mei\ss ner-Institut, Bayerische Akademie der Wissenschaften, D-85748 Garching, Germany}
\author{R. Gross}
\affiliation{Walther-Mei\ss ner-Institut, Bayerische Akademie der Wissenschaften, D-85748 Garching, Germany}
\affiliation{Physik-Department, Technische Universit\"at M\"unchen, D-85748 Garching, Germany}
\affiliation{Nanosystems Initiative Munich (NIM), Schellingstra\ss e 4, 80799 M\"unchen, Germany}
\author{E. Solano}
\affiliation{Department of Physical Chemistry, University of the Basque Country UPV/EHU, Apartado   644, 48080 Bilbao, Spain}
\affiliation{IKERBASQUE, Basque Foundation for Science, Maria Diaz de Haro 3, 48013 Bilbao, Spain}

\begin{abstract}
Propagating quantum microwaves have been proposed and successfully implemented to generate entanglement, thereby establishing a promising platform for the realisation of a quantum communication channel. However, the implementation of quantum teleportation with photons in the microwave regime is still absent. At the same time, recent developments in the field show that this key protocol could be feasible with current technology, which would pave the way to boost the field of microwave quantum communication. Here, we discuss the feasibility of a possible implementation of microwave quantum teleportation in a realistic scenario with losses. Furthermore, we propose how to implement quantum repeaters in the microwave regime without using photodetection, a key prerequisite to achieve long distance entanglement distribution.
\end{abstract}

\maketitle

\section{Introduction}
In 1993, C.H. Bennett et al.~\cite{Bennett1} proposed a protocol to disassemble a quantum state at one location (Alice) and to reconstruct it in a spatially separated location (Bob). They proved that, if Alice and Bob share quantum correlations of EPR type~\cite{Einstein1}, then Bob can reconstruct the state of Alice by using classical channels and local operations. This phenomenon is called ``quantum teleportation'', and it has important applications in quantum communication~\cite{Ralph13}. The result inspired discussions among physicists, in particular, on the experimental feasibility of the protocol. Despite some controversies in technical issues, the first experimental realisation of quantum teleportation was simultaneously performed in 1997 in two groups, one led by A. Zeilinger in Innsbruck~\cite{Zeilinger1}, and the other by F. De Martini in Rome~\cite{Boschi1}. In both experiments, the polarisation degrees of freedoms of the photons were teleported. It was shown that, even within the unavoidable experimental errors, the overlap between the input state and the teleported one exceeded the classical threshold achievable when quantum correlations are not present. After the success of the first experiments, alternatives for a variety of systems and degrees of freedom emerged. Of particular interest is the continuous-variable scheme studied by L. Vaidman~\cite{Vaidman1} and S. L. Braunstein et al.~\cite{Braunstein1}, whose experimental implementation was realised by A. Furusawa et al.~\cite{Furusawa1} in the optical regime. This experiment consisted in teleporting the information embedded in the continuous values of the conjugated variables of a propagating electromagnetic signal in the optical regime. Optical frequencies were preferred because of their higher detection efficiency, essential to achieve a high fidelity performance~\cite{Braunstein1, Furusawa1}, and because propagation losses are almost negligible. During the last years, an impressive progress in teleporting quantum optical states to larger distances, first in fibers~\cite{Marcikic1, Ursin1}, and afterwards in free-space~\cite{Jin1, Yin1, Ma1}, was made. This rapid progress may even allow us to realise quantum communication via satellites in near future with corresponding distances of about $150$ km. In optical systems, the long-distance teleportation is, to some extent, straightforward, because of the high transmissivity of optical photons in the atmosphere. Nevertheless, unavoidable losses are setting an upper limit for the teleportation distance. However, there were fundamental theoretical studies on how to allow for a long-distance entanglement distribution. The underlying concepts are based on quantum repeaters~\cite{Briegel1, Duan1}, whose implementation on specific platforms needs an individual study.  So far, the entanglement sharing and quantum teleportation was reported for cold atoms~\cite{Riebe1, Barrett1, Olmschenk1}, and even for macroscopic systems~\cite{Bao1}.

In this article, we discuss the possibility of implementing the quantum teleportation protocol of propagating electromagnetic quantum signals in the microwave regime. This line of research is justified by the recent achievements of circuit quantum electrodynamics (cQED)~\cite{Blais1, Wallraff1}. In cQED, a quantum bit (qubit) is implemented using the quantum degrees of freedom of a macroscopic superconducting circuit operated at low temperatures, i.e. $\laq$ $10$-$100$ mK, in order to suppress thermal fluctuations. Superconducting Josephson junctions are used to introduce non-linearities in these circuits, which are essential in both quantum computation and the engineering of qubits. Typical qubits are built to have a transition frequency in the range $5$-$15$ GHz (microwave regime), and they are coupled to an electromagnetic field with the same frequency. This choice is determined by readily available microwave devices and techniques for this frequency band, such as low noise cryogenic amplifiers, down converters, network analysers, among others. We note that  apart from its relevance in quantum communication, quantum teleportation is also crucial to perform quantum computation, e.g. it can be used to build a deterministic CNOT gate~\cite{Gottesman}. 

Recently, path-entanglement between propagating quantum microwaves has been investigated in Refs.~\cite{Menzel1, Eichler1, Flurin1}. Following what was previously done in the optical regime~\cite{Ou1}, a two-mode squeezed state, in which the modes were spatially separated from each other, was generated. The two entangled beams could be used to perform with microwaves a protocol equivalent to the one used in optical quantum teleportation~\cite{Braunstein1, Furusawa1}. These articles represent the most recent of a large amount of results presented during the last years~\cite{Menzel2, Mariantoni1, Mallet1, Bozyigit1, Eichler1, Eichler2, Eichler3, Eichler4, daSilva1, DiCandia1, Zhong1, Yamamoto1, Castellano1, Hoffmann1, Mariantoni2, Baur1, Lanzagorta1}, which are the building blocks of a quantum microwave communication theory.
Inspired by the last theoretical and experimental results, we want to discuss the feasibility of a quantum teleportation realisation for propagating quantum microwaves. The article is organised in the following way: In Section~II, we introduce the continuous-variable quantum teleportation protocol and its figures of merit. In Section~III, we describe the preparation of a propagating quantum microwave EPR state. In Section~IV, we show how to implement a microwave equivalent of an optical homodyne detection, by using only linear devices. The Section~V is focused on the analysis of losses. In particular, we consider an asymmetric case in which the losses in Alice's and Bob's paths are different. In Section~VI, we discuss the feedforward part of the protocol in both a digital and an analog fashion. Finally, as the entanglement distribution step is affected by losses, we present in Section~VII how to implement a quantum repeater based on weak measurements in a cQED setup, in order to allow the entanglement sharing at larger distances.

\section{The Protocol}

In this Section, we briefly explain the quantum teleportation protocol introduced in~\cite{Vaidman1, Braunstein1}, and we introduce some useful figures of merit to quantify the quality of the scheme in a realistic setup. The protocol consists in teleporting a continuous variable state, and it has already been applied in the optical regime to both a Gaussian state~\cite{Furusawa1, Takei1} and a Schr\"odinger cat state~\cite{Lee1}. An equivalent scheme for microwaves is still missing, therefore a specific treatment, in which the restriction imposed by the technology  is taken into account, is mandatory to analyse its feasibility. Let us consider a situation in which two parties, Alice and Bob, want to share a quantum state. More specifically, Alice, labelled with $A$, wants to send a quantum state $|\phi\rangle_T$, whose corresponding system is labeled  by $T$, to Bob, denoted by $B$. Additionally, let them share an ancillary entangled state $|\psi\rangle_{AB}$ given by
\begin{equation}\label{EPR}
(\hat{x}_A+\hat{x}_B) | \psi\rangle_{AB}= \delta(x_A+x_B),\quad\quad (\hat{p}_A-\hat{p}_B) |\psi\rangle_{AB}= \delta(p_A-p_B),
\end{equation}
where $\hat x$ and $\hat p$ are quantum conjugate observables obeying the standard commutation rule $[\hat x,\hat p]=i$. After Alice performs a Bell-type measurement on the system $T$-$A$,
\begin{equation}\label{measEPR}
 x_T+ x_A=a\quad\quad  p_T- p_A=b,
\end{equation}
where $a$ and $b$ are the outcomes of the measurement. The resulting values of Bob's quadrature would be
\begin{equation}\label{telep}
 x_B= x_T-a,\quad\quad  p_B= p_T-b.
\end{equation}
By displacing adaptively Bob's state by $a+ib$, i.e. $x_B$ is shifted by $a$ and $p_B$ by $b$, we finally have $\hat x_B|\phi\rangle_B=\hat x_T| \phi\rangle_T$ and $\hat p_B |\phi\rangle_B=\hat p_T| \phi\rangle_T$, where $ |\phi\rangle_B$ is the final state of Bob. Therefore, the final state of Bob is the state of the system $T$. Note that Bob needs to perform local operations conditioned to Alice's measurement outcomes. As the outcomes are two numbers, we may allow Alice and Bob to communicate throughout a classical channel, see Fig.~1. Bennett et al.~\cite{Bennett1} called this protocol a \emph{quantum teleportation}~\cite{Peres1}.

A state fulfilling~\eqref{EPR} can be seen as a two-mode squeezed state with infinite squeezing. In fact, its Wigner function can be written as
\begin{align}
W_{\text{A-B}}(x_A,p_A,x_B,p_B)&=\frac{1}{\pi^2} \exp \left\{-\frac{e^{-2r}}{2}\left[(x_A-x_B)^2+(p_A+p_B)^2\right]-\frac{e^{+2r}}{2}\left[(x_A+x_B)^2+(p_A-p_B)^2\right]\right\}  \nonumber\\
\quad&\sim\; \frac{2}{\pi e^{2r}}\exp \left\{-\frac{e^{-2r}}{2}\left[(x_A-x_B)^2+(p_A+p_B)^2\right] \right\}\delta(x_A+x_B)\delta(p_A-p_B), \label{WignerEPR}
\end{align}
where $r$ is a squeezing parameter~\cite{Walls1} and, also, we have considered an asymptotic behaviour for large $r$. For finite $r$, the state of the system $A$-$B$ fullfils 
\begin{equation}
\hat{x}_A+\hat{x}_B=\hat{\xi}_x,\quad\quad \hat{p}_A-\hat{p}_B=\hat{\xi}_p,
\end{equation}
where $\hat \xi_x|\psi\rangle_{AB}$ and $\hat \xi_p|\psi\rangle_{AB}$ have real Gaussian distributions with mean value equal to zero and variance $e^{-2r}$. If we perform the teleportation protocol with this state, the final Wigner function for Bob's state is the weighted integral
\begin{equation}\label{Wigner1}
W_B(x_B,p_B)=\int d\xi_x d\xi_p  P(\xi_x)P(\xi_p) W_T(x_B-\xi_x,p_B+\xi_p),
\end{equation}
where $P(\xi_{x,p})$ are the probability distributions of the outcomes of $\hat \xi_{x,p}$. After introducing the variables $x_B-\xi_x=X$, $p_B+\xi_p=Y$, and defining $\alpha=X+iY$, $z_B=x_B+ip_B$, we get
\begin{equation}\label{conv}
W_B(z_B)=\int d^2\alpha P_c(z_B^*-\alpha^*) W_T(\alpha) 
\end{equation}
where $P_c$ is the complex Gaussian distribution with mean value zero and variance $\bar{\sigma}^2=e^{-2r}$, i.e. $P_c(\beta)=\frac{1}{2\pi\bar{\sigma}^2}\exp\left\{\frac{-|\beta|^2}{2\bar{\sigma}^2}\right\}$. In the limit of infinite $r$, $P_c$ approaches to the delta function, and then $W_B=W_T$. In the following, we will refer only to the variance of the quadratures, regardless of whether they are noisy or not. Therefore, our treatment is general, and it includes also the lossy case, in which we do not have a perfect two-mode squeezed state as a resource. In order to evaluate the performance of the protocol, entanglement fidelity~\cite{Schumacher1} can be used. If $T$ is in a pure state, the entanglement fidelity is given by
\begin{equation}
\mathcal{F}=\pi\int dz_B^2 dz_T^2 W_B(z_B)W_T(z_T).
\end{equation}
If Alice is restricted to teleport coherent states, the protocol works better than in the classical case corresponding to $r=0$ if $\mathcal{F}>\frac{1}{2}$ \cite{Braunstein2}.
Let us remark that the performance of the protocol for coherent states, and in general for Gaussian states, depends only on the variances $\Delta  \xi_{x,p}^2\equiv \langle \Delta \hat \xi_{x,p}^2\rangle$. Indeed, one can verify that \cite{Takei1}
\begin{equation}
\mathcal{F}=\frac{1}{\sqrt{(1+\Delta \xi_x^2)(1+\Delta \xi_p^2)}}, 
\end{equation}
 and $\mathcal F>\frac{1}{2}$ is valid if and only if
\begin{equation}\label{csi}
\Xi\equiv(1+\Delta  \xi_x^2)(1+\Delta  \xi_p^2)<4.
\end{equation}
More general cases could also be discussed, but this does not provide any additionally insight into the question under which conditions the protocol is feasible. The condition in Eq.~\eqref{csi} defines our limit between classical and quantum teleportation. While in the noiseless case this condition is satisfied for any positive squeezing, the situation changes when we take losses into account. From now on, we assume the case of coherent state teleportation, and the symmetric case, where $\Delta \xi^2\equiv\Delta \xi^2_{x,p}$ and $\Delta \xi_\perp^2\equiv \langle\Delta (\hat x_A-\hat x_B)^2\rangle=\langle\Delta (\hat p_A+\hat p_B)^2\rangle$.

\begin{figure}[t]
\centering
\includegraphics[width=0.90\textwidth]{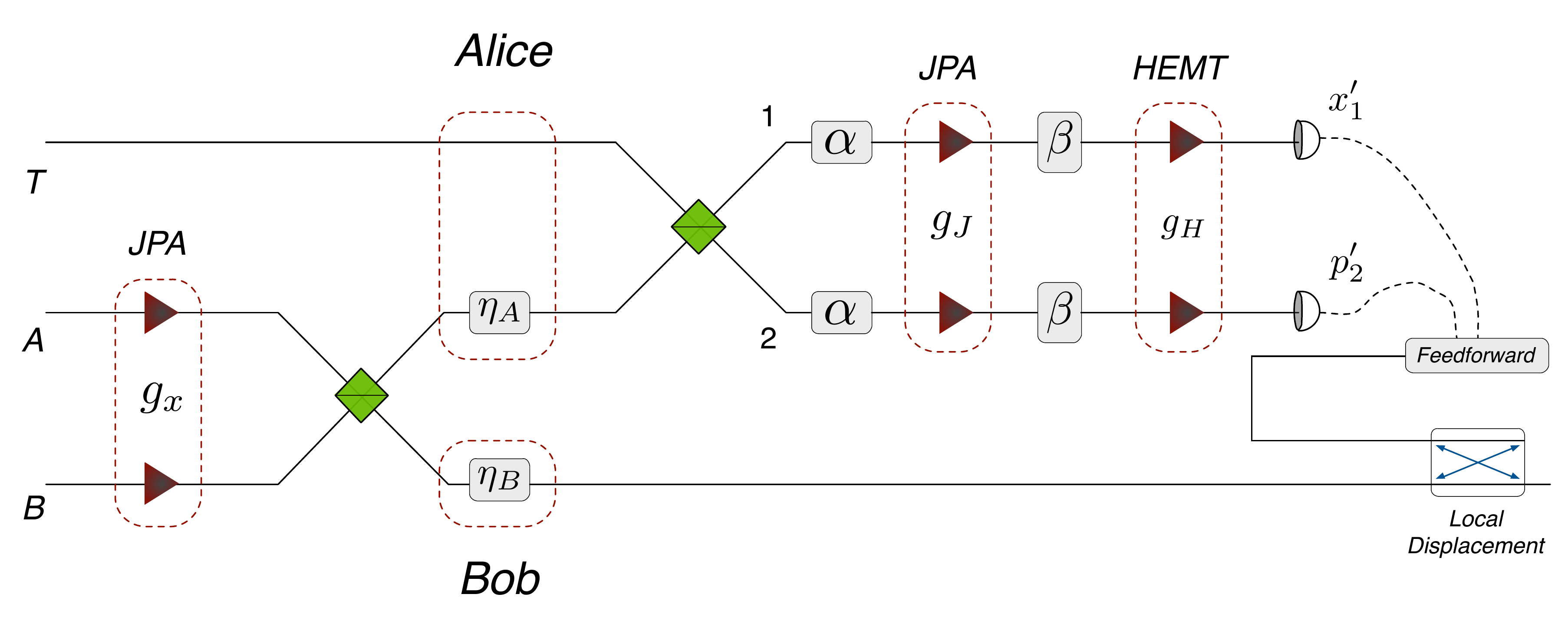}\label{figtel}
\caption*{}
\begin{flushleft}
{\bf Fig. 1}. Scheme of the proposed quantum teleportation protocol. The generation of an EPR state is obtained by amplifying the orthogonal vacuum quadratures of $A$ and $B$ with two Josephson parametric amplifiers (JPAs). The generated entanglement  is then shared between Alice and Bob. Alice uses this resource to perform a Bell-type measurement with the state that she wants to teleport. This is realised by superposing her two signals with a beam splitter, and then measuring a quadrature in each of the outputs. The quadrature measurement is performed via amplifying the signal with a JPA and a HEMT amplifier in series, and then measuring via homodyne detection. Finally, after a classical transfer of Alice results, a local displacement on the Bob state is needed to conclude the protocol. The figure indicates where losses (labeled as $\eta_{A,B}^{}$, $\alpha$, $\beta$)  may be present.
\end{flushleft}
\end{figure}

\section{Generation of EPR State}
Following Refs.~~\cite{Menzel1, Flurin1}, propagating quantum microwave EPR states are prepared in the following way. We can generate a microwave vacuum state with a $50$ Ohm resistor at low temperatures $T\sim50$ mK, as its blackbody radiation corresponds to a thermal state with number of photons $n_\omega=(e^{ \hbar\omega/kT}-1)^{-1}$, with $n_\omega\ll1$ for frequencies $\omega/2\pi\sim5$-$15$ GHz. By sending the vacuum to a Josephson parametric amplifier (JPA)~\cite{Yamamoto1, Castellano1}, we can create a one-mode squeezed state, in which the squeezed quadrature is defined by the phase of the JPA pump signal. The relation between the input $\hat a_{in}^{}$ and the output $\hat a_{out}^{}$ of a JPA~\cite{Caves1} 
\begin{equation}\label{JPA}
\hat a_{out}^{}=\hat a_{in}^{}\cosh r+\hat a_{in}^\dag \sinh r,
\end{equation} 
is the same as for a squeezing operator. Notice that the amplified quadrature is defined by $\hat x_{out}^{}=(\hat a_{out}^{}+\hat a_{out}^{\dag})/\sqrt{2}$, and the squeezed quadrature is the orthogonal one. A two-mode squeezed state~\cite{Walls1} can be generated by sending two one-mode squeezed states, squeezed with respect to orthogonal quadratures, to a hybrid ring, acting as a microwave beam splitter \cite{Hoffmann1, Mariantoni2}. In this way, the resulting Wigner function is given by Eq.~\eqref{WignerEPR}, and the two output modes are spatially separated (see Fig.~1). 

\begin{figure}[t]
\centering
\includegraphics[width=0.60\textwidth]{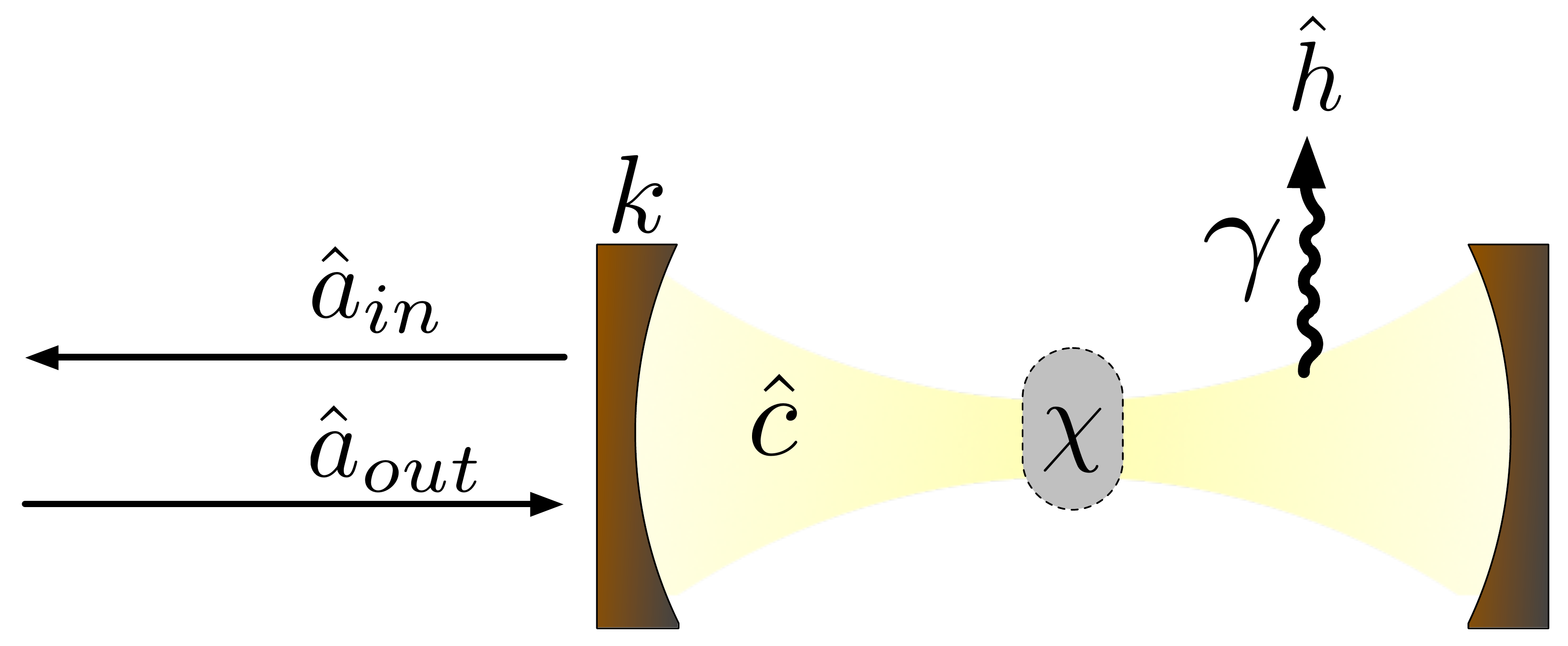}\label{paramp}
\vspace{0.5cm}
\caption*{}
\begin{flushleft}
{\bf Fig. 2}. Scheme of a Josephson parametric amplifier (JPA). The field outside the resonator interacts with the resonator mode with a coupling rate $k$. The resonator mode is evolving under the squeezing Hamiltonian with a coupling $\chi$. The losses are taken into account by introducing an environment mode $\hat h$, and let it interact with the resonator mode with a coupling rate $\gamma$. 
\end{flushleft}
\end{figure} 

In general, the quality of the entanglement between the two modes is affected by the losses of the JPA. To take into account the inefficiency, we write down the Hamiltonian of the JPA and take into account a finite coupling of the resonator mode $\hat c$ with an environment, as depicted in Fig.~2:
\begin{align}\label{eq1}
H=H_{free}+i\frac{\hbar \chi}{2}(c^2 -c^{\dag 2}) +i\hbar \sqrt{\frac{k}{2\pi}} \int d\omega [a(\omega)c^\dag-ca^\dag(\omega)]+i\hbar \sqrt{\frac{\gamma}{2\pi}} \int d\omega [h(\omega)c^\dag-ch^\dag(\omega)],
\end{align}
where $H_{free}=\hbar \omega_c c^\dag c+\int d\omega\; \hbar \omega \;a^\dag(\omega) a(\omega)+\int d\omega\; \hbar \omega \;h^\dag(\omega)h(\omega)$ is the free Hamiltonian. The second term in Eq.~\eqref{eq1} is the squeezing Hamiltonian, the third term models the interaction between the cavity field and the input and output signals, and the last term takes into account the losses. The output mode of the JPA is defined as the steady state of $\hat a$. One can write down these equations in the Heisenberg picture, and look at input-output relations of the fields:
\begin{align}
\hat x_{a_{out}}&=\frac{2\chi+k-\gamma}{2\chi-k-\gamma}\hat x_{a_{in}}+\frac{2\sqrt{k\gamma}}{2\chi-k-\gamma}\hat x_{h_{in}}\equiv \sqrt{g_x}\; \hat x_{a_{in}}+\sqrt{s_x}\; \hat x_{h_{in} \label{amplx}},\\
\hat p_{a_{out}}&=\frac{2\chi-k+\gamma}{2\chi+k+\gamma}\hat p_{a_{in}}-\frac{2\sqrt{k\gamma}}{2\chi+k+\gamma}\hat p_{h_{in}}\equiv \frac{1}{\sqrt{g_p}}\;\hat p_{a_{in}} -\sqrt{s_p}\; \hat p_{h_{in}}, \label{amplp}
\end{align}
where the $h_{in}$ label refers to the input noise, assumed to be a thermal state fulfilling the relation $s_xs_p=\left(\sqrt{g_x/g_p}-1\right)^2$\\~\cite{Caves1}. 
The quantities $\Delta \xi^2$ and $\Delta \xi^2_\perp$ introduced at the end of the Section~II can be easily retrieved by using Eqs.~\eqref{amplx}-\eqref{amplp} and the beam splitter relation:
\begin{align}
\Delta \xi^2= \frac{1}{g_p}+2s_p\;\Delta p_{h_{in}}^2\quad \Delta \xi_\perp^2= g_x+2s_x\;\Delta x_{h_{in}}^2.
\end{align} 
Note that $\gamma=0$ corresponds to a noiseless parametric amplifier, whose input-output relations are shown in Eq.~\eqref{JPA}, with $e^{r}\equiv\sqrt{g_x}=\sqrt{g_p}$. Generally, the JPA generates a squeezed thermal state whose squeezed quadrature has variance $\sigma_s^2$. We have entanglement between the outputs of the hybrid ring if $\sigma_s^2<\sigma_{vac}^2$, where $\sigma_{vac}^2\equiv0.5$ is the variance of the vacuum. 
The variance measured in~\cite{Menzel1} is $\sigma_s^2\simeq 0.16$, which leads, considering a beam splitter with $0.4$~dB of power losses, to an EPR state with $\Delta \xi^2\simeq0.47$ ($\Delta\xi_\perp^2\simeq 16.77$) and $\Xi\simeq 1.74<4$. In the following, we will use these values as reference, altough we believe that these parameters can be improved with better JPA designs. 

\section{Quadrature Measurement}
Measuring a quadrature of a weak microwave signal is considered a particularly difficult task, since the low energy of microwave photons makes it difficult to realise a single-photon detector. Therefore, the standard homodyne detection scheme is not applicable. Typically, one has to amplify the microwave signal in order to detect it. Cryogenic high electronic mobility transistor (HEMT) amplifiers are routinely used in quantum microwave experiments~\cite{Menzel1, Menzel2, Mallet1, Bozyigit1, Eichler1, Eichler2, Eichler3, Mariantoni2}, because of their large gains in a relatively broad frequency band. However, HEMT amplifiers are phase insensitive and add a significant amount of noise photons, sufficient to make the quantum teleportation protocol fail. Their input-output relations are~\cite{Caves1}
\begin{equation}\label{HEMT}
\hat a_{out}=\sqrt{g_H}\;\hat a_{in}+\sqrt{g_H-1}\;\hat h_H^\dag,
\end{equation}
where $\hat a_{in}$, $\hat a_{out}$ and $\hat h_H$ are annihilation operators of the input field, output field and noise added by the amplifier, respectively, with $g_H\sim10^4$ for modern high-performance cryogenic amplifiers. We can assume $\hat h_H^{}$ to be in a thermal state with thermal population $n_H$. For instance, commercial cryogenic HEMT amplifiers have a typical number of added noise photons $n_H\sim 10$-$100$ for the considered frequency regime.\\
To measure $\hat x_T+\hat x_A$ and $\hat p_T-\hat p_A$, we need to send the state A and the state T to a hybrid ring, obtaining
\begin{align}
\hat a_{1}=\frac{\hat a_T+\hat a_A}{\sqrt{2}}\quad\quad \hat a_2=\frac{\hat a_T-\hat a_A}{\sqrt{2}},
\end{align}
and then measure the $x$-quadrature of the mode $1$ and the $p$-quadrature of the mode $2$. If we amplify this signal with a HEMT and then measure it afterwards, the state of Bob after the local displacement is
\begin{align}
\hat x_B= \hat x_T+\hat \xi_x+\sqrt{\frac{2(g_H-1)}{g_H}}\hat x_{h_H},
\end{align}
and analogously for $\hat p_B$. One can easily check that even if the added noise photons are at the vacuum level, we get $\mathcal{F} \leq \frac{1}{2}$ and the protocol fails. 

To avoid this situation, we can adopt a scheme based on anti-squeezing the target quadrature before the HEMT amplification~\cite{Leonhardt1, Mallet1}, see Fig.~1. Corresponding outputs of the amplification JPAs with a gain $g_J$, followed by a HEMT amplification with gain $g_H$, are   
\begin{align}
\hat  x'_{1}&=\sqrt{g_Hg_J}\;\hat x_{1}^{}+\sqrt{g_Hs}\;\hat x_{h_{J}}+\sqrt{g_H-1}\;\hat x_{h_{H}}^{},
\end{align}
and similar for $\hat p'_{2}$. We assume, for the sake of simplicity,  the symmetric case, where both quadratures have the same amplification, and the amount of added noise is the same in both modes. The state of Bob after the displacement step is
\begin{align}
\hat x_B^{}= \hat x_T^{}+\hat \xi_x^{}+\sqrt{\frac{2(g_H-1)}{g_Jg_H}}\hat x_{h_{H}}^{}+\sqrt{\frac{2s}{g_J}}\hat x_{h_{J}}^{},
\end{align}
and analogously for $\hat p_B$. In the limit of large $g_J$, the noise of the HEMT amplifier is suppressed and the inefficiencies of the JPA are negligible, provided that as  $\Delta x_{h_{J}}^2$ and $s$ are small. By defining the JPA quadrature noise $A_{J}\equiv\frac{s}{g_J}\;\Delta \hat x_{h_J}^2$, and the HEMT quadrature noise $ A_{H}\equiv\frac{g_H-1}{g_H}\Delta \hat x_{h_H}^2$,
we have to analyse for which experimental values the total noise
\begin{equation}
A\equiv 2\left(A_{J}+\frac{A_{H}}{g_J}\right)
\end{equation}
is lowest, since for $A>1$ the protocol fails. In the recent experiments on quantum state tomography of itinerant squeezed microwave states~\cite{Mallet1}, an additional JPA with a degenerate gain $g_J^{}\simeq180$ was used as a preamplifier. Corresponding figures of merit are $A_{J}\simeq0.25$, and in case of $A_{H}\simeq17$, we get $A\simeq0.69$. With these values, if we take into account the quality of the EPR state mentioned at the end of Section~III, the protocol fails, as $\Xi=(1+\Delta \xi^2+A)^2\simeq4.04>4$. However, the HEMT quadrature noise can realistically reach a value of $A_H\simeq 7$, and this gives us an upper bound to the JPA quadrature noise in order for the quantum teleportation protocol to work, i.e. $A_J<0.30$. This bound does not take into account losses and measurement inefficiencies, which are considered in the next Section. Moreover, we believe that JPA values can certainly be improved within the next years, as JPA technology is considerably advancing both in the design and  materials~\cite{Mutus1, Eom1, Brien1, Eichler2011, Macklin2015}.

\section{Protocol with Losses}
So far, we have not taken into account possible losses in the protocol. Typically, losses in the microwave domain are much larger than in the optical domain, and therefore can significantly affect the quality of the teleportation protocol. In the following, we analyse the protocol with all possible loss mechanisms, see Fig.~1. Note that losses after the HEMT amplification are negligible, and therefore omitted.

\begin{figure}[t]
\centering
\includegraphics[width=\textwidth]{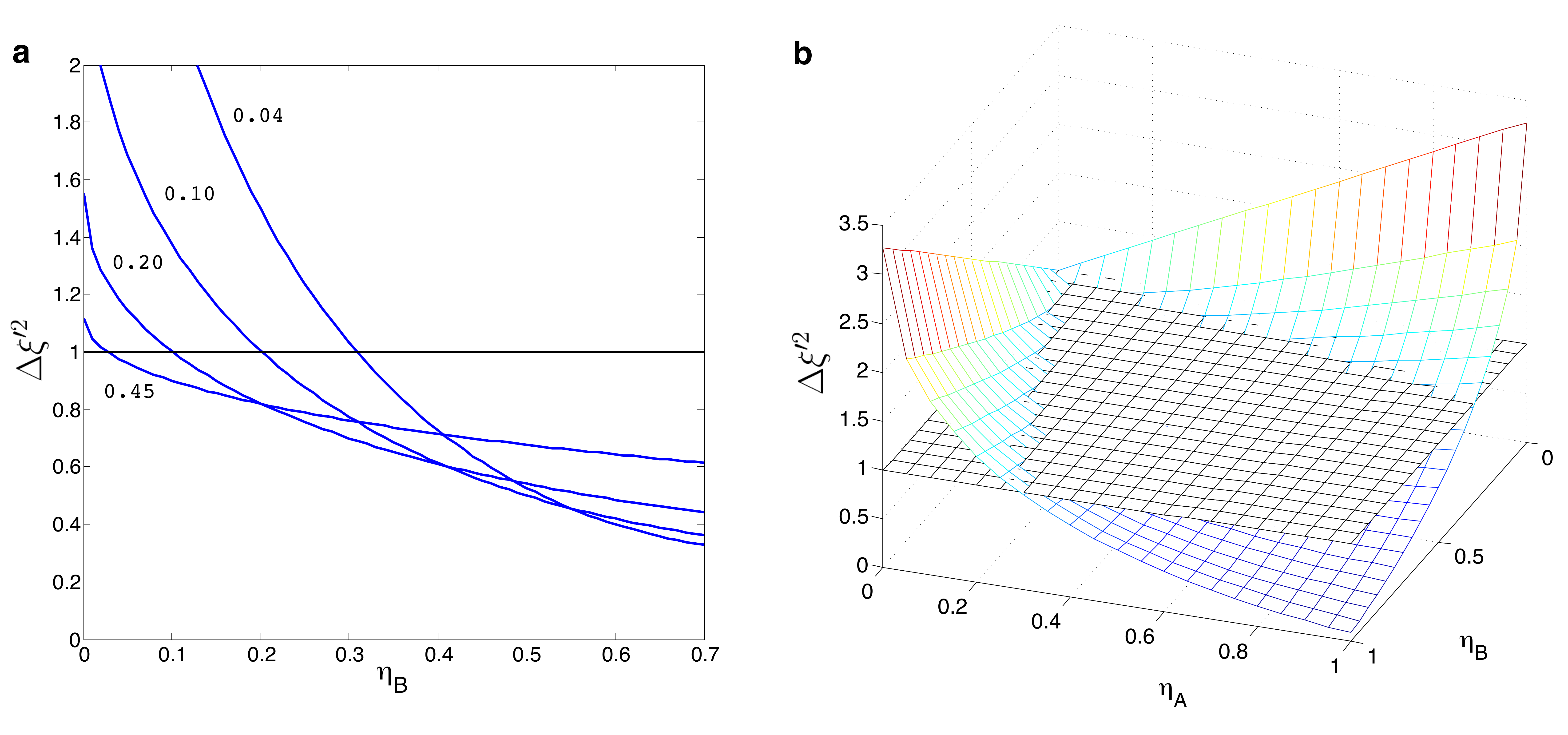}\label{Figure2}
\caption*{}
\begin{flushleft}
{\bf Fig. 3}. The quantity to $\Delta \xi'^2$ defined in Eq.~\eqref{deltax}, which describes the amount of correlations between Alice and Bob, plotted as a function of the transmission coefficients $\eta_{A,B}$ modelling the losses in Alice's and Bob's channel. The case $\Delta \xi'^2\geq 1$ corresponds to a classically reachable performance. We see that the quality of the protocol depends on a compromise between squeezing, given by $\Delta \xi^2$, and transmissivity coefficients, given by $\eta_{A,B}$. {\bf a}. $\Delta \hat \xi'^2$ plotted as a function of $\eta_B$ for fixed $\eta_A=0.70$ and  for various values of $\Delta \xi^2$, assuming $\eta_B<\eta_A$ and noiseless EPR-JPAs. $\Delta \xi^2$ determines the entanglement in the lossless case: the entanglement increases with decreasing $\Delta \xi^2$. We see that the window of the allowed difference between the losses in Alice's and Bob's channel reduces for larger entanglement. {\bf b}. Here, $\Delta \hat \xi'^2$ is plotted as function of $\eta_B$ and $\eta_A$ for fixed $\Delta \xi^2=0.14$. From Eq.~\eqref{deltax}, we see that for a too large asymmetry between Alice's and Bob's channel, it is opportune to symmetrize them by attenuating one of the signals in order to increase the amount of correlations between the two parties.  For instance, for $\eta_B=0.3$ and $0.8<\eta_A<1$, we find that $\Delta \hat \xi'^2$ increases with increasing $\eta_A$.
\end{flushleft}
\end{figure}

To characterise the losses, we use a beam splitter model. Following Fig.~1, the fields after collecting the losses in the entanglement sharing step are
\begin{equation}
\hat x'_A=\sqrt{\eta_A}\,\hat x_A+\sqrt{1-\eta_A}\,\hat x_{v_A}
\quad
\hat x'_B=\sqrt{\eta_B}\,\hat x_B+\sqrt{1-\eta_B}\,\hat x_{v_B},
\end{equation}
where $\eta_{A,B}$ are the transmission coefficients modelling the losses in Alice's and Bob's channel respectively, and $\hat x_{v_{A,B}}$ are modes in a thermal state (similar formulas hold for $\hat p_{A,B}$). Then,
\begin{align}
\hat x'_A+\hat x'_B&=\frac{\sqrt{\eta_A}+\sqrt{\eta_B}}{2}(\hat x_A+\hat x_B)+\frac{\sqrt{\eta_A}-\sqrt{\eta_B}}{2}(\hat x_A-\hat x_B)  +\sqrt{1-\eta_A}\,\hat x_{v_A}+\sqrt{1-\eta_B}\,\hat x_{v_B}\equiv \hat  \xi' ,
\end{align}
and
\begin{align} \label{deltax}
\Delta {\xi'}^2&=\frac{(\sqrt{\eta_A}+\sqrt{\eta_B})^2}{4}\Delta \xi^2+\frac{(\sqrt{\eta_A}-\sqrt{\eta_B})^2}{4}\Delta \xi_\perp^2 +(1-\eta_A)\left(n_{v_A}+\frac{1}{2}\right)+(1-\eta_B)\left(n_{v_B}+\frac{1}{2}\right).
\end{align}
We note that the second term in Eq.~\eqref{deltax} results from an asymmetry of the losses in Alice's and Bob's channel and it increases with squeezing level in the EPR JPAs.
In the optical domain, $\eta\sim1$, allowing to neglect this term even for asymmetric channels. Moreover, in this frequency range, $n_{v_{A,B}}\ll1$ even at room temperature. In the microwave domain, instead, we have $n_{v_{A,B}}\sim10^3$ at room temperature and typical power losses of $20\%$ per meter. In this case, the entanglement would collapse after $\sim 2$ mm regardless of the value of $g_x$. Thus, in the following we assume that the entanglement distribution is possible at $50$ mK, i.e. $n_{v_{A,B}}\ll1$. As already pointed out, if $\eta_A\not=\eta_B$, then $\Delta \xi'^2$ contains a term linearly increasing with the JPA gain $g_x$. Equation~\eqref{deltax} explains why the ideal quantum teleportation, i.e. $\mathcal{F}=1$, is not possible in a realistic experiment even with in the limit of infinite squeezing as input. From Fig.~3a, we see that the allowed difference between $\eta_A$ and $\eta_B$ decreases with decreasing $\Delta \xi^2_{}$. In Fig.~3b, instead, we see that for large differences between $\eta_A$ and $\eta_B$, it is convenient to attenuate the signal of Alice. For instance, if $\eta_B<\eta_A$, we can easily see that this happens when $\frac{\partial \Delta \xi'^2}{\partial \eta_A}>0$, i.e.
\begin{equation}\label{ineq}
\sqrt{\frac{\eta_B}{\eta_A}}<\frac{\left(\Delta \xi_{}^2+\Delta \xi_\perp^2\right)/4-\left(n_{v_A}+\frac{1}{2}\right)}{\left(\Delta \xi_\perp^2-\Delta \xi_{}^2\right)/4}.
\end{equation}
As Alice's measurement step takes a finite amount of time, we typically have $\eta_B<\eta_A$.

Concerning Alice's measurement, we may define the quantities characterising the noise added by losses as
\begin{align}
A_{\alpha}^{}\equiv\frac{1-\alpha}{\alpha}\Delta  x_{v_\alpha}^{2},\quad A_{\beta}^{}\equiv\frac{1-\beta}{\beta}\Delta  x_{v_\beta}^{2}.
\end{align}
Here, $\alpha$ is the transmission coefficient from the the output of the hybrid ring to the JPA, taking into account the hybrid ring losses. Moreover, $\beta$ is the transmission coefficient from the JPA to HEMT amplifier. Hence, the total noise is
\begin{equation}\label{totalnoise}
A=2\left(A_{\alpha}^{}+\frac{A_J^{}}{\alpha}+\frac{A_{\beta}}{\alpha g_J^{}}+\frac{A_H}{\alpha\beta  g_J^{}}\right),
\end{equation}
where $A_J$ and $A_H$ were defined in the previous Section. 

In Table~1, we estimate a bound on $A_J$ for typical losses, taking into account the feedforward (discussed in the following Section), and for several distances. These numbers imply that the device experimentally investigated in Refs.~\cite{Zhong1, Mallet1}, two of the few available studies of JPA noise in the degenerate mode, are already close to the threshold where a benefit over classical approaches can be achieved. We immediately see that the largest contributions to $\Xi$ come from $A_J$ and $\Delta \xi'^2$. For example, a version of the protocol would work if the noise added by the detection amplifiers is reduced by a good factor of three to $A_J<0.073$, corresponding to $1$~m distance from the EPR source. Similarly, improvements in the EPR state generation would help via a reduced $\Delta \xi'^2$. Regarding the latter, particular attention should be given to the distance over which an EPR pair can be distributed. For our numbers, assuming a superconducting coaxial cable of $1$~m length, the dominating contributions to the losses still come from the beam splitter and connectors. Therefore, an implementation of our protocol for the quantum microwave communication between two adjacent chips of a superconducting quantum processor or two superconducting quantum information units in nearby buildings seems feasible with some reasonable technological improvements. In this context, we want to reiterate that the big advantage of the quantum microwave teleportation lies in the fact that microwaves are the natural operating frequencies of superconducting quantum circuits.

\begin{figure}[t]
\centering
\includegraphics[width=0.7\textwidth]{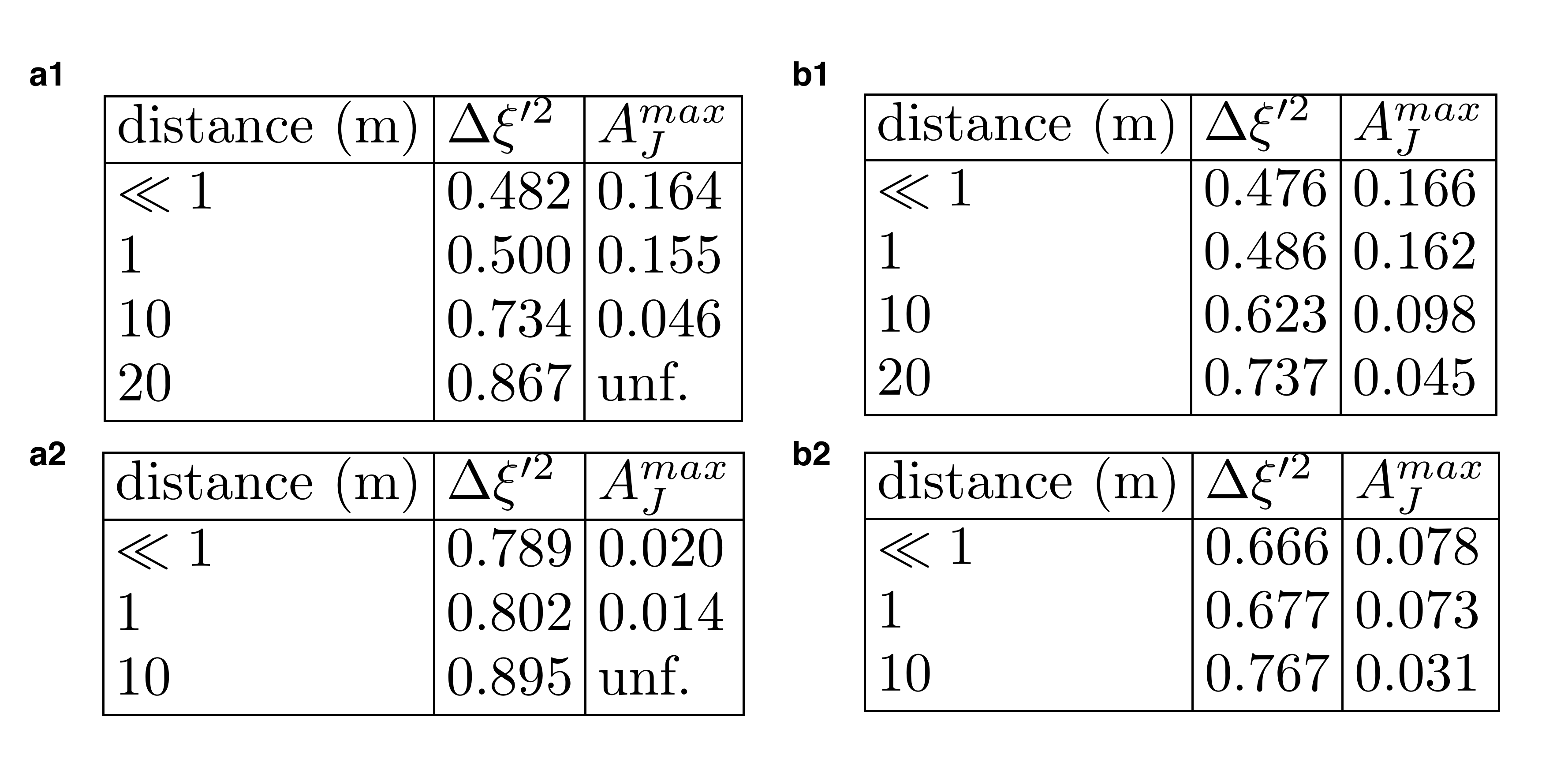}\label{table}
\caption*{}
\begin{flushleft}
{\bf Table 1}. Tables with the maximum value of $A_J^{max}$ allowed in order for the quantum teleportation protocol to work. The abbreviation ``unf.'' means ``unfeasible". We assume an EPR state with the values discussed in Section~III, i.e. $\Delta \xi^2\simeq 0.47$, $\Delta \xi_\perp^2\simeq 16.77$, and typical values for connector losses leading to $\alpha\simeq0.933$ ($A_\alpha\simeq 0.036$), $\beta\simeq0.891$ ($A_\beta\simeq 0.061$). Moreover, we assume a JPA gain $g_J\simeq 180$, and HEMT noise $A_H \simeq 7$. The noise parameter $A_J^{max}$ is estimated from the formulas $\Xi\equiv (1+\Delta \xi'^2 +A)^2\leq 4$, which defines the quantum regime. Here, $\Delta \xi'^2$ is defined in Eq.~\eqref{deltax} and $A$ is introduced in Eq.~\eqref{totalnoise}. We assume Alice and Bob symmetrically situated with respect the EPR sources. The distance is referred to the cable length from the EPR sources to Alice (Bob). The estimations take into account of the feedforward, and $A_J^{max}$ is evaluated for various distances and in four different situations. In {\bf a1}. we assume cable power losses of $0.1$~dB per meter and zero time measurement. In {\bf a2}  we assume cable power losses of $0.1$~ dB per meter and $200$~ns for measuring and processing the information in Alice. These two tables give an insight on how much the measurement duration, which result in a delay line in Bob, affects the quality of the protocol. In {\bf b1} we assume a more optimistic value for cable power losses, i.e. $0.05$~dB per meter, and zero time measurement. In {\bf b2} we assume $0.05$~dB of power losses per meter and $200$~ns for measuring the processing the information in Alice. In all the tables, when Eq.~\eqref{ineq} holds, we have applied a proper attenuator in Alice in order to optimise $\Delta \xi'^2$.
\end{flushleft}
\end{figure}

\section{Analog vs. Digital Feedforward}

In the quantum teleportation protocol, Alice needs to measure and send the result of the measurement to Bob via a {\itshape classical} channel. Then, Bob uses this information to apply a displacement in his system. This process is called a feedforward, and is considered  tough to implement, independently of the considered system.  In particular, in the microwave case, the measurement process may be slow, resulting in an ultimate loss of fidelity. In realistic experiments, a quantum microwave signal has to be amplified before detection. If the amplification is large, the signal becomes insensitive to losses at room temperature. Therefore, an idea is to use the output signal of Alice to perform classical communication without digitally measuring it. This analog feedforward is depicted in Fig.~4, and it works in the following way. Let us assume the lossless case, and send the two amplified signals of Alice to a hybrid ring. One of the two outputs of the latter provides us with 
\begin{align}
\hat x_F^{}&=\frac{1}{2}\left(\sqrt{g_Jg_H}-\sqrt{\frac{g_H}{g_J}}\right)\hat x_A^{}+\frac{1}{2}\left(\sqrt{g_Jg_H}+\sqrt{\frac{g_H}{g_J}}\right)\hat x_T^{}+\sqrt{\frac{g_H-1}{2}}(\hat x_{h_{H1}}^{}+\hat x_{h_{H2}}^{}) ,\\
\hat p_F^{}&=\frac{1}{2}\left(-\sqrt{g_Jg_H}+\sqrt{\frac{g_H}{g_J}}\right)\hat p_A^{}+\frac{1}{2}\left(\sqrt{g_Jg_H}+\sqrt{\frac{g_H}{g_J}}\right)\hat p_T^{}-\sqrt{\frac{g_H-1}{2}}(\hat p_{h_{H1}}^{}+\hat p_{h_{H2}}^{}),
\end{align}
where the label ``F'' stands for the feedforward. Indeed, Bob may use this signal to perform the displacement. 

\begin{figure}[t]
\centering
\includegraphics[width=0.80\textwidth]{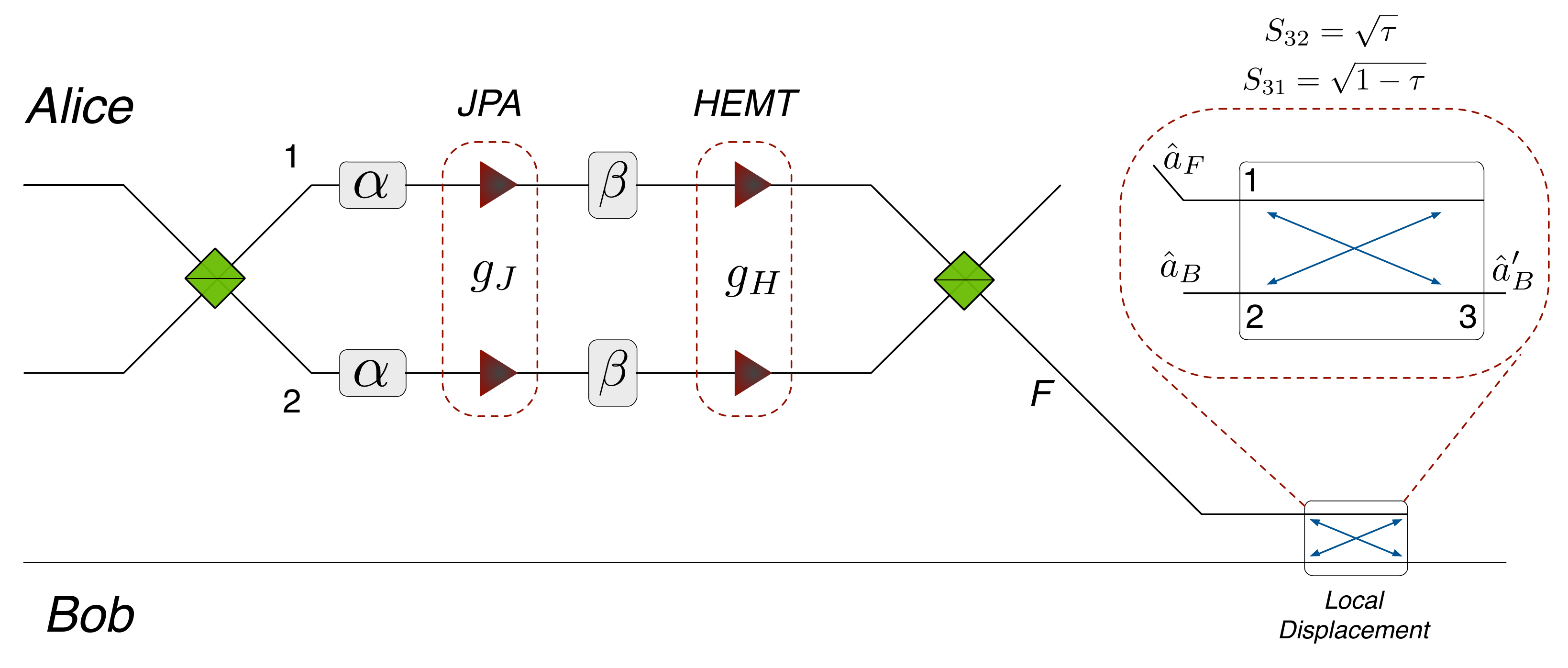}\label{analogfeed}
\caption*{}
\begin{flushleft}
{\bf Fig. 4}. Scheme of the analog feedforward. Here, Alice is not digitising the signals, but  is amplifying and superposing them. The output signal is robust to the environment noise, and it contains all the information that Bob needs to perform the local displacement. This displacement is then implemented with a high transmissivity directional coupler, whose inputs are the signal of Bob, and the output signal of Alice. In the figure, $\hat a_F$ is the output signal of Alice, $\hat a_B$ is the signal of Bob, $\hat a'_B$ is the output of the teleportation scheme, while $\tau\simeq 1$ is the transmissivity and $S_{ij}$ the scattering matrix of the directional coupler.
\end{flushleft}
\end{figure}
A displacement operator can be implemented by sending a strong coherent state and the field which we want to displace to a high-transmissivity mirror \cite{Paris1}. Hence, the transmitted signal is 
\begin{equation}\label{displacement}
\hat a_{out}=\sqrt{\tau}\,\hat a_{in}+\sqrt{1-\tau}\,  \alpha,
\end{equation}\label{displacement1}
where $\alpha$ is without a hat because it represents a coherent state. If we choose $\tau\sim1$ and $|\alpha|\gg1$ such that $\sqrt{1-\tau}\,\alpha=z$, we obtain
\begin{equation}\label{displacement}
\hat a_{out}=\sqrt{\tau}\,\hat a_{in}+z\;\simeq\; \hat a_{in}+z,
\end{equation}
which approximates a displacement operator. In a microwave experiment, the operation \eqref{displacement} can be implemented with a microwave directional coupler. If we send signals B and F as inputs to a directional coupler with transmissivity $\tau\simeq1-\frac{4}{g_Jg_H}$, the corresponding output is 
\begin{align}
\hat x_B'&=\sqrt{\tau}\,\hat x_B+\sqrt{1-\tau}\,\hat x_F=\left(1+\frac{1}{g_J}\right)\hat x_T+\sqrt{1-\frac{4}{g_Jg_H}}\hat x_B+\left(1-\frac{1}{g_J}\right)\hat x_A + \nonumber\\
\quad&+\sqrt{\frac{2(g_H-1)}{g_Jg_H}}(\hat x_{h_{H1}}+\hat x_{h_{H2}}) \simeq \hat x_T+\hat \xi_x , \label{directional1}\\
\hat p_B'&=\sqrt{\tau}\,\hat p_B+\sqrt{1-\tau}\,\hat p_F=\left(1+\frac{1}{g_J}\right)\hat p_T+\sqrt{1-\frac{4}{g_Jg_H}}\hat p_B+\left(\frac{1}{g_J}-1\right)\hat p_A - \nonumber\\
\quad&-\sqrt{\frac{2(g_H-1)}{g_Jg_H}}(\hat p_{h_{H1}}+\hat p_{h_{H2}})\simeq \hat p_T-\hat \xi_p,\label{directional2}
\end{align}
where the last approximation holds for $g_J\gg1$, and, for the sake of simplicity, we have considered the lossless case.  Considering the typical values $g_H\sim10^4$ and $g_J\sim10^2$, we would need a reflectivity factor $1-\tau\sim10^{-6}$. For this value, small errors in $\tau$ would result in a large error in the displacement operator. This problem can be overcome by attenuating at low temperatures the signal F before the directional coupler, in order to neglect the attenuator noise. In this case, setting $\tau=1-\frac{4}{\eta_{att}g_Jg_H}$, the transmitted signal is the same as in \eqref{directional1}-\eqref{directional2}. For instance, if we choose $\eta_{\text{att}}\sim10^{-3}$, we derive a reasonable value for the reflectivity: $1-\tau\sim10^{-3}$.

The described analog method allows us to perform the feedforward without an actual knowledge of the result of Alice's measurement. Indeed, the JPA and HEMT amplifiers work as measurement devices. On the one hand, the advantage is that we save the time required to digitalised the signal. On the other hand, the disadvantage is that all the noise sources in Alice are mixed, resulting in a doubling of the noise $A$, as we see in Eqs.~\eqref{directional1}-\eqref{directional2} (the same claim holds for the lossy case). Therefore, one should carefully evaluate whether the digital feedback is convenient against the analog one, by comparing $A$, which quantify the loss of fidelity in the analog feedforward case, with the noise added due to the delay line added in Bob in the digital feedforward case. This can be done by estimating the digitisation time and the corresponding losses in the Bob delay line, which strongly depends on the available technology.  Indeed, currently available IQ mixers and FPGA technology requires $t_p\sim 200-400\;$~ns for measuring and processing the information. During this time, the signal needs to be delayed in Bob's channel. If we consider a delay line where the group velocity of the electromagnetic field is $v\simeq 2 \times 10^8\;$~m/s, $t_p$ corresponds to a delay line in Bob of $40-80$~m. Comparing the values of $\Delta \xi'^2$ for the zero measurement time and the realistic $200$~ns measurement time, we see a change in $\Delta \xi'^2$ of $\sim0.30$ in the case of $1$~m distance (assuming $0.1$~dB per meter of power cable losses), which is considerably lower than the current values achievable for $A$. Notice that this discrepancy decreases with the distance between Alice and Bob. This means that the digital feedforward is currently preferable to the analog one, but the analog feedforward can become a useful technological tool when the JPA technology will reach a reasonable noise level.

\section{Quantum Repeaters}

As we have discussed in the previous Sections, the entanglement distribution between the two parties, Alice and Bob, is particularly challenging due to the large losses involved. Moreover, while in the optical case the noise added by a room temperature environment corresponds to the vacuum, in the microwave regime, this noise would correspond to a thermal state containing $\sim10^3$ photons. Even in the most favourable situation in which we build a cryogenic setup to share the entanglement, we would have a collapse of the correlations after $\sim 10$ m due to the detection inefficiency and losses. The implementation of quantum repeaters in the microwave regime could potentially solve this issue. A quantum repeater is able to distillate entanglement and to share it at larger distance, at the expense of efficiency. A protocol for distributing entanglement at large distance in the microwave regime has been recently proposed in~\cite{Tombesi}, but it relies on the implementation of an optical-to-microwave quantum interface~\cite{Tombesi2}, which has not yet been realised experimentally. 

Here, we discuss the microwave implementation of quantum repeater based on a non-deterministic noiseless linear amplification via weak measurements~\cite{Menzies}. A noiseless linear amplifier~\cite{Ralph, Xiang} can be modeled as an operator $g^{\hat n}$ applied to its input state. For example, for a input coherent state $|\alpha\rangle$, we would have $|g\alpha\rangle$ as output, resulting in a amplification of all quadratures without adding noise. Let us consider a two-mode squeezed state $|\psi_{AB}\rangle\propto \sum_{n=0}^{\infty}(\tanh r)^n|n\rangle_A|n\rangle_B$. Notice that the amount of entanglement increases on increasing $r$. If we are able to implement the operator $g^{\hat n}$, with $g>1$ on one mode, say Bob, we have  
\begin{align}
g^{\hat n_B}|\psi_{AB}\rangle\propto \sum_{n=0}^{\infty}(g \lambda)^n|n\rangle_A|n\rangle_B=\sum_{n=0}^{\infty}\lambda'^n|n\rangle_A|n\rangle_B,
\end{align}
with $\lambda=\tanh r$ and $\lambda'\equiv g \lambda >\lambda$. A similar argument holds, if we have losses in each of the two modes. In fact, the state after the loss mechanism is 
\begin{align}
|\psi_{loss}\rangle&\propto \sum_{n=0}^\infty\sum_{k_A=0}^n\lambda^n\sum_{k_B=0}^n(-1)^{2n-k_A-k_B}\eta_A^{k_A/2}\eta_B^{k_B/2}(1-\eta_A)^{(n-k_A)/2}\nonumber\\
\;&(1-\eta_B)^{(n-k_B)/2}\sqrt{\binom{n}{k_A}\binom{n}{k_B}}|k_A\rangle_A|k_B\rangle_B|n-k_A\rangle_{l_A}|n-k_B\rangle_{l_B},
\end{align}
where $l_{A,B}$ correspond to the loss modes. If we apply the operator $g^{\hat n_B}$, the output state has the same form but with the new effective parameters~\cite{Ralph} $\eta_B\rightarrow \eta_B'=\frac{g^2\eta_B}{1+(g^2-1)\eta_B}$ and $\lambda\rightarrow \lambda''=\lambda\sqrt{1+(g^2-1)\eta_B}$, which is accompanied by an increase of the entanglement. Accordingly, the final $\Delta \xi'^2$ would be lower, which corresponds to higher values of $A_J^{max}$ in Table~1.
Note that if $\lambda=0$, i.e. no entanglement at the input, then the output state is not entangled either. Therefore, in order to increase the amount of entanglement, we need a minimum of entanglement at the input. 

The operator $g^{\hat n}$ corresponds to a noiseless phase-insensitive linear amplifier, and it cannot be implemented deterministically. However, there exist probabilistic methods to realise it approximately. A probabilistic noiseless linear amplification scheme has already been demonstrated in the optical regime~\cite{Ralph,Xiang}, but it relies on the possibility of counting photons. In contrast, the weak measurement scheme~\cite{Menzies} requires quadrature measurements that can be applied in the microwave regime.

Let Bob's mode interact with an ancillary system in a coherent state $|\alpha\rangle$ accordingly to the cross-Kerr Hamiltonian {$\hat H_{\text{Kerr}}= \hbar k\;\hat n_{anc}\hat n_B$}, where $k$ is a coupling constant. Let us further consider low-time interaction, i.e. $k\Delta t\ll1$. If we postselect the ancilla in the state $|p\rangle$, i.e. the eigenstate of the $\hat p$ quadrature corresponding to the eigenvalue $p$, the whole final state is
\begin{align}\label{weak}
|\psi_{final}\rangle&=|p\rangle\langle p|\,e^{-i\hat H_{\text{Kerr}}\Delta t/\hbar}|\alpha\rangle|\psi\rangle_{AB}\simeq |p\rangle\langle p|\,(\mathbb{I}-ik\Delta t\;\hat n_{anc}\hat n_B)|\alpha\rangle|\psi\rangle_{AB} \nonumber\\
\;&=|p\rangle\langle p|\alpha\rangle\,\left(\mathbb{I} -ik\Delta t A_{w}\hat n_B\right)|\psi\rangle_{AB}\simeq|p\rangle\langle p|\alpha\rangle\, e^{-ik\Delta tA_{w}\hat n_B }|\psi\rangle_{AB} \nonumber\\
\;&=|p\rangle\langle p|\alpha\rangle\, e^{-ik\Delta t\,{\rm Re} (A_w)\hat n_B }\left(e^{k\Delta t \,{\rm Im}(A_w)}\right)^{\hat n_B}|\psi\rangle_{AB}
\end{align}
where $A_w\equiv\frac{\langle p| \hat n_{anc}|\alpha\rangle}{\langle p|\alpha\rangle}=\alpha^2-i\sqrt{2}\alpha p$ is called ``weak value'', and, in the second approximation, we have assumed {$k\Delta t|A_{w}|\ll1$}. By choosing appropriately the values of $\alpha$ and $p$, we can induce a value of $A_w$, whose imaginary part is positive. If we set $g\equiv e^{k\Delta t\,{\rm Im} (A_w)}$, we have a scheme to implement $g^{\hat n_B}$ up to a known phase-shift $e^{-ik\Delta t\,{\rm Re} (A_w)\hat n_B }$, with success probability density $|\langle p|\alpha\rangle|^2=\frac{1}{\sqrt{\pi}}e^{-\left(p-{\rm Im}(\alpha)\right)^2}$. For instance, by choosing ${\rm Im}(\alpha)=0$ and ${\rm Re}(\alpha) <0$, we have a gain for any $p>0$, which happens with a $50\%$ probability. In this case, an imperfect quadrature measurement can be corrected by just shifting the allowed results of the ancilla measurement, with a consequent lost of efficiency. Note that, due to the probabilistic nature of the scheme, Alice and Bob need to communicate classically in order to distillate the entanglement, see Fig.~5. However, this classical communication can be performed at the end, in a post-selection fashion, as Alice does not need to perform any operation on her system.

\begin{figure}[t]
\centering
\includegraphics[width=0.80\textwidth]{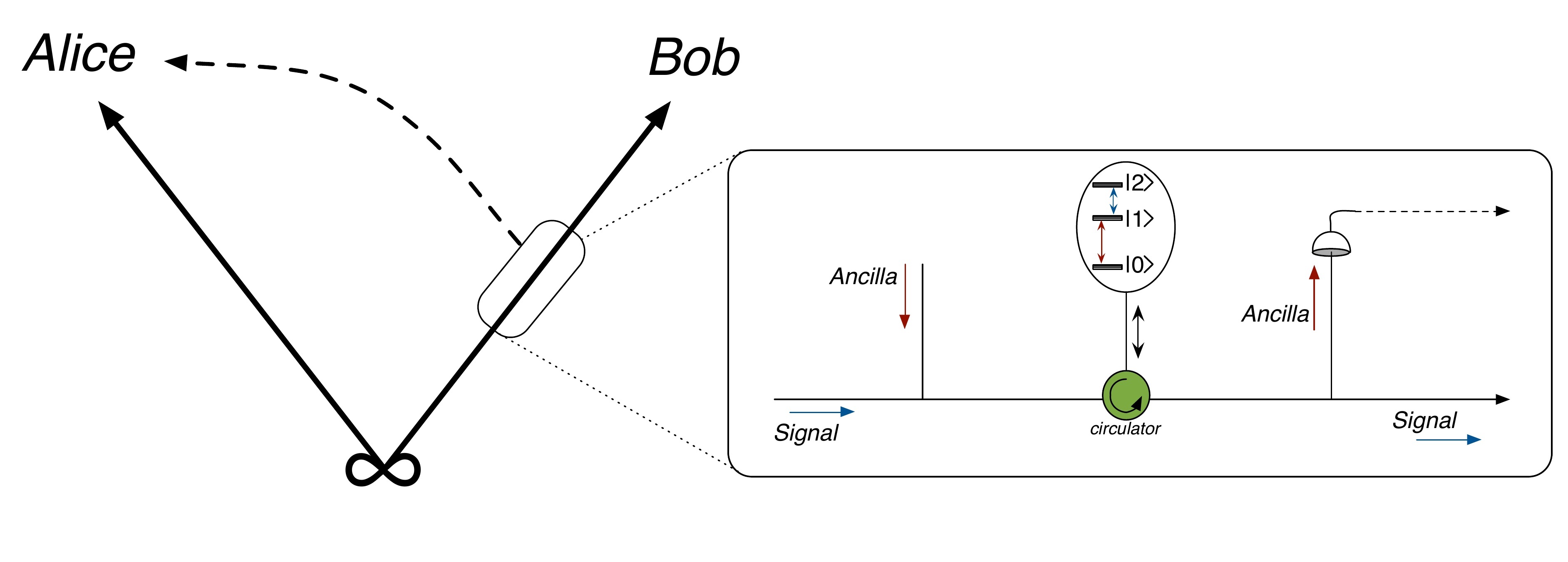}\label{qrepeater}
\caption*{}
\begin{flushleft}
{\bf Fig. 5}. Quantum repeater scheme with weak measurement and postselection. A probabilistic noiseless linear amplifier is applied to one of the two parties, via the implementation of a weak cross-Kerr interaction with an ancillary signal. This interaction emerges as a fourth order expansion of the dynamics of the signal and the ancilla coupled with a transmon quantum bit, modeled as a three level system, in a non-resonant regime. The ancilla is then measured, and the result is sent classically to Alice for post-selection.
\end{flushleft}
\end{figure}

The cross-Kerr effect, characterised by a Hamiltonian of the kind $\hat H_{\text{Kerr}}=\hbar k\;\hat n_{anc}\hat n_B$, has already been proposed in cQED in the context of single-photon resolved photodetectors, see~\cite{Hoi, Sathy}. Basically, this interaction emerges in the fourth order expansion of the dynamics of two microwave modes coupled with a transmon in a non-resonant regime. By modelling the transmon as a three-level system, the system Hamiltonian is
\begin{align}
H=\hbar \omega_a a^\dag a +\hbar \omega_b b^\dag b + \hbar (\omega_2-\omega_0)|2\rangle\langle 2| + \hbar (\omega_1-\omega_0)|1\rangle\langle 1| +\hbar g_a\left[a|2\rangle\langle1| +a^\dag |1\rangle\langle2|\right]+\hbar g_b\left[b|1\rangle\langle0| +b^\dag |0\rangle\langle1|\right],
\end{align}
where $a$ represents the ancillary mode and $b$ Bob's mode.
In the interaction picture with respect $H_0=\hbar \omega_a a^\dag a+\hbar \omega_b b^\dag b+\hbar \omega_a |2\rangle\langle 2|+\hbar \omega_b|1\rangle\langle1| $, the new Hamiltonian is
\begin{align}
H_I=\hbar \Delta_a |2\rangle\langle2| +\hbar \Delta_b |1\rangle\langle1| +\hbar g_a\left[a|2\rangle\langle1| +a^\dag |1\rangle\langle2|\right]+\hbar g_b\left[b|1\rangle\langle0| +b^\dag |0\rangle\langle1|\right],
\end{align}
where $\Delta_a=\omega_2-\omega_a$, $\Delta_b=\omega_1-\omega_b$, and we have set $\omega_0=0$. If we set the parameters in order to have $g_a,g_b\ll \Delta_a,\Delta_b,|\Delta_a-\Delta_b|$, and we inizialize the transmon in $|0\rangle$, the effective Hamiltonian is
\begin{align}\label{Jeff}
H_I^{eff}=\hbar \frac{g_b^2}{\Delta_b}b^\dag b\; |0\rangle\langle0| +\hbar \frac{12g_a^2g_b^2}{\Delta_a\Delta_b}\left(\frac{1}{\Delta_b}-\frac{1}{\Delta_a}\right)a^\dag a b^\dag b \;|0\rangle\langle 0|,
\end{align}
where we have implemented a fourth order expansion of the Magnus series, and we have used the rotating wave approximation. Typical parameters allowing this are $(\omega_2-\omega_1)/2\pi=\omega_a/2\pi\simeq 5$~GHz, $(\omega_1-\omega_0)/2\pi=(\omega_b+\tilde \Delta)/2\pi$, with $\tilde \Delta =20$~MHz and $\omega_b/2\pi\simeq 6$~GHz, and $g_{a,b}\simeq 100$~kHz. The Hamiltonian in Eq.~\eqref{Jeff} represents the cross-Kerr effect up to a known phase, that can be corrected at the end. In this scheme, dissipations are negligible, as we are interested in very low interaction times.

\section{Conclusions}

We have considered a quantum teleportation protocol of propagating quantum microwaves. We have analysed its realisation by introducing figures of merit (i.e. $\Xi$ and $A$) that takes into account losses and detector efficiency. In particular, we have underlined the difference between the optical case (where photodetectors are available, and losses are negligible) and the microwave regime. Indeed, we have considered JPAs in order to perform single-shot quadrature measurements, and we have proposed an analog feedforward scheme, which does not rely on digitisation of signals. Moreover, we have discussed the losses mechanisms, highlighting in which measure they limit the realisation of the protocol. We have used typical parameters of present state-of-art experimental setups in order to identify the required improvements of these setups to allow for a first proof-of-principle experiment. Finally, we have introduced a quantum repeater scheme based on weak measurements and postselection.

\section*{Acknowledgments}
This work is supported by the German Research Foundation through SFB 631, and the grant FE 1564/1-1; Spanish MINECO FIS2012-36673-C03-02; UPV/EHU UFI 11/55; Basque Government IT472-10; CCQED, PROMISCE, and SCALEQIT EU projects.

\section*{ADDITIONAL INFORMATION} 
The authors declare no competing financial interests.

\end{document}